\documentstyle[mprocl]{article}
\def\Journal#1#2#3#4{{#1} {\bf #2}, #3 (#4)}

\def\be{\begin{equation}}
\def\ee{\end{equation}}
\def\bea{\begin{eqnarray}}
\def\eea{\end{eqnarray}}

\begin{document}

\title{GLOBALLY HYPERBOLIC GEODESICALLY COMPLETE COSMOLOGICAL MODEL}

\author{ Leonardo Fern\'andez Jambrina }

\address{Dep. Ense\~nanzas B\'asicas, \\ETSI Navales, \\Avenida Arco de la Victoria
s/n, \\E-28040-Madrid, Spain}

\maketitle\abstracts{
In this talk we shall show a perfect fluid cosmological model and its 
properties. The model possesses an orthogonally transitive abelian
two-dimensional group of isometries that corresponds to cylindrical symmetry.
The matter content is a stiff fluid that satisfies the energy and generic
conditions. The metric is not separable in comoving coordinates for the fluid. 
The curvature invariants are shown to be regular everywhere in the coordinate
chart and also indicate that the spacetime is asymptotically flat.
Furthermore the causal geodesics are studied in order to determine
that they are complete and that the model is globally hyperbolic.
The model goes through an initial contracting epoch that is
followed by an expanding era.}
  
\section{History}

In 1990 it was published the first known non-singular cosmological solution  of
the Einstein equations for a perfect fluid with physically reasonable
properties \cite{Seno}. Later on it was proven that this solution was
geodesically complete and globally hyperbolic as well as the absence of
causally trapped sets \cite{Chinea}. These results triggered research on
regular cosmological exact solutions. 

A larger family of separable diagonal
orthogonally transitive commuting $G_2$ metrics  was found \cite{Ruiz} and
included in a general metric with  FLRW cosmologies  \cite{family}.
Furthermore non-diagonal non-singular cosmological models have been shown
 \cite{Diag}$^{\!,\,}$ \cite{Jerry}$^{\!,\,}$ \cite{Mars} and also non-separable 
regular solutions can be found in the literature \cite{Sep}. 

\section{Another non-singular model}

In this talk a cosmological model with cylindrical symmetry for a
stiff perfect fluid is presented. It is non-singular and it is
non-separable in comoving coordinates. Its metric tensor,

\begin{equation}\label{metric}
ds^2=-\theta^0\otimes\theta^0+\theta^1\otimes\theta^1+\theta^2\otimes\theta^2+
\theta^3\otimes\theta^3,
\ee
which has been written in terms of an orthonormal coframe,
\begin{equation}
\theta^0={\rm e}^{\frac{1}{2}\,K(t,r)}\,dt,\ \ \ \ \theta^1={\rm
e}^{\frac{1}{2}\,K(t,r)}\,dr, \ \ \ \ \theta^2={\rm e}^{-\frac{1}{2}
U(t,r)}\,dz, \ \ \ \ \theta^3=
 {\rm e}^{\frac{1}{2}U(t,r)}\,r\,d\phi,\label{coframe}
\end{equation}

\begin{equation}
K(t,r)=\frac{1}{2}\,{ \beta}^{2}\,{r}^{4 } + ( \alpha+\beta)\,{r}^{2} +
2\,{t}^{2}\,{ \beta}
 + 4\,{t}^{2}\,{ \beta}^{2}\,{r}^{2},
 \end{equation}
 \begin{equation}
 U(t,r)={ \beta}\,(\,{r}^{2} + 2\,{t}^{2}\,), \ \ \ \alpha,\beta\ge 0
 \end{equation}
has been expressed in a chart where the coordinates $\phi$ and $z$ are 
adapted to the Killing fields and the parametrization of the metric on the
subspaces orthogonal to the group orbits is isothermal. The time coordinate
ranges from minus infinity to infinity. The other coordinates have the usual
range for cylindrical symmetry.

 The metric is generically Petrov type I and does not admit further
non-trivial isometries. The gradient of the transitivity
surface element is spacelike as in every other non-singular model.

 The expressions for the pressure and the density of the fluid, 

\begin{equation}
p=\mu= \alpha\,{\rm e}^{-K(t,r)},
\end{equation}
 show that the Ricci curvature scalars are regular everywhere in the chart.

In the natural  orthonormal coframe the kinematical quantities of the fluid read

\be
u=\theta^0, \ \ \ a={r}\, \left( \! \,{ \beta}^{2}\,{r
}^{2} + { \alpha} + { \beta} + 4\,{ \beta}^{2}\,{t}^{2}\, \! 
 \right) \, {\rm e}^{\frac{1}{2}K(t,r)}\,\theta^1,\ \ \ \omega=0,
\ee

\begin{eqnarray}
\sigma=\frac{4}{3}\,{ \beta
}\,{t}\,{\rm 
e}^{-\frac{1}{2}\,K(t,r)}\,\left\{(\,1 + 2\,{ 
\beta}\,{r}^{2}\,)\,\theta^1\otimes \theta^1-\right.\\\nonumber\left.-(\,2 + { 
\beta}\,{r}^{2}\,)\,\theta^2\otimes \theta^2+(1-\,{ \beta}\,{r}^{2
})\,\theta^3\otimes 
\theta^3\right\},
\end{eqnarray}

\begin{equation}
{ \Theta}=2\,\beta\,t\,(\,1 + 2\,{ \beta}\,{r}^{2}\,)\,{\rm 
e}^{-\frac{1}{2}\,K(t,r)}.
\end{equation}

The deceleration parameter for this cosmology \cite{Ellis} can be calculated
from the expansion of the fluid. It shows that this universe undergoes a finite
inflationary phase from 
$-\tau$ to $\tau$,
\begin{equation}
\tau(r)= \left[{\frac {3}{4\,{ \beta}\,(1 + 2\,{
 \beta}\,{r}^{2})}}\right]^\frac{1}{2},
\end{equation}
as it happens in other non-singular cosmological models.

\section{Physical properties of the model}

It has already been shown that the Ricci curvature scalars are regular.
Furthermore  all the curvature scalars are regular since it can be checked
that the components of the Weyl tensor in a complex orthonormal coframe
\cite{sing} are bounded in the chart. They tend to zero for large
values of the time and radial coordinates. Hence the universe is asymptotically
Minkowski spacetime. The universe contracts until
$t=0$ and then begins to expand.

The matter content of the universe is a stiff perfect fluid with positive
density everywhere in the chart. This means that both the strong and dominant
energy conditions \cite{HE} and the generic condition \cite{Beem} are fulfilled.

Since the most common definition of a spacetime singularity that is found in the
literature is the existence of a causal geodesic which is incomplete, it will be
necessary to study the geodesic equations for this model. The work is
simplified by the existence of two independent constants of motion, which are
related to translations along the $z$-axis and rotations around it.

The system of five second order geodesic equations in the affine
parametrization can be shown to reduce to three first order
equations plus two quadratures \cite{sing}. From this system of equations
a reasoning can be devised to establish that the range of the affine
parameter of every causal geodesic is the real line. Therefore the spacetime is
causally geodesically complete.

Concerning the causality conditions it is easy to notice that the model is
causally stable since the coordinate function $t$ is a cosmic time \cite{HE}.
From the null geodesic completeness a stronger condition, global hyperbolicity,
follows.

According to the singularity theorems \cite{HE}$^{\!,\,}$
\cite{Beem} this spacetime cannot contain trapped sets. 

\section*{Acknowledgments}
The present work has been supported by Direcci\'on General de
Ense\~nanza Superior Project PB95-0371 and by a DAAD (Deutscher Akademischer
Austauschdienst) grant for foreign scientists.  The author wishes to thank
Prof. F. J. Chinea and Dr. L. M. Gonz\'alez-Romero for valuable discussions
and Prof. Dietrich Kramer and the Theoretisch-Physikalisches Institut of the 
Friedrich-Schiller-Universit\"at-Jena for their hospitality.

\end{document}